\documentclass[12pt]{article}

\usepackage{natbib}
\usepackage{graphicx}
\usepackage{amssymb}
\usepackage{color}
\usepackage{setspace}

\newcommand{\bea}{\begin{eqnarray}}
\newcommand{\eea}{\end{eqnarray}}
\newcommand{\be}{\begin{equation}}
\newcommand{\ee}{\end{equation}}

\newcommand{\ldphage}{$\lambda$}

\newcommand{\Mg}{\rm{Mg^{+2}}}

\definecolor{Mygreen}{cmyk}{0,1.0,0,0}
\definecolor{MyOrange}{cmyk}{0.5,1.0,0,0}

\newcommand{\Rin}{R_{\rm{in}}}
\newcommand{\Rout}{R_{\rm{out}}}
\newcommand{\Ucapsid}{U_{\rm{capsid}}}
\newcommand{\Uvesicle}{U_{\rm{vesicle}}}
\newcommand{\Osmotic}{\Pi_{\rm{osmotic}}}
\newcommand{\Fosmotic}{F_{\rm{osmotic}}}
\newcommand{\RDNA}{R_{\rm{DNA}}}
\newcommand{\tauo}{\tau_{\rm 0}}
\newcommand{\tauocc}{\tau_{\rm occ}}
\newcommand{\tauun}{\tau_{\rm unocc}}
\begin{document}


\title{Dynamics of DNA Ejection From Bacteriophage}
\author{Mandar M. Inamdar$^{*}$,  William M. Gelbart$^{\#}$, and Rob
  Phillips$^{*,\S,\dagger}$}
\maketitle

\noindent $^{*}$Division of Engineering and Applied Science,
      California Institute of Technology, Pasadena, CA 91125.\\
\noindent $^{\S}$ Kavli Nanoscience Institute, California Institute of
Technology, Pasadena, CA 91125. \\
\noindent $^{\#}$Department of Chemistry and Biochemistry, University
of California, Los Angeles, CA 90024. \\
\noindent $^{\dagger}$ To whom the correspondence should be addressed: phillips@pboc.caltech.edu

\begin{abstract}
The ejection of DNA from a bacterial virus (``phage'') into its
host cell is a biologically important example of the translocation
of a macromolecular chain along its length through a membrane. The
simplest mechanism for this motion is diffusion, but in the case
of phage ejection a significant driving force derives from the
high degree of stress to which the DNA is subjected in the viral
capsid. The translocation is further sped up by the ratcheting and
entropic forces associated with proteins that bind to the viral
DNA in the host cell cytoplasm.  We formulate a generalized
diffusion equation that includes these various pushing and pulling
effects and make estimates of the corresponding speed-ups in the
overall translocation process. Stress in the capsid is the
dominant factor throughout early ejection, with the pull due to
binding particles taking over at later stages. Confinement effects
are also investigated, in the case where the phage injects its DNA
into a volume comparable to the capsid size. Our results suggest a
series of \textit{in vitro} experiments involving the ejection of
DNA into vesicles filled with varying amounts of binding proteins
from phage whose state of stress is controlled by ambient salt
conditions or by tuning genome length.
\end{abstract}



\section{\label{sec:intro}Introduction}

A crucial first step in the life cycle of most bacterial viruses
involves binding of the virion to a receptor protein in the host cell
membrane followed by injection of viral DNA. The viral genome is
typically about 10 microns long, and its translocation from outside to
inside the host cell is accomplished over times that vary from seconds
to minutes. The wide range of mechanisms responsible for injection of
phage genomes has recently been systematically
reviewed~\citep{molineux01, letellier04, gonzales-Huici04}, including
many references to the last few decades of relevant literature. In the
present paper we formulate a general theory of chain translocation
that takes into account many of the physical processes involved in
actual phage life cycles. These processes include: diffusion of the DNA chain
along its length; driving forces due to stress on the DNA inside the viral
capsid; resisting forces associated with osmotic pressure in the host
cell; cell confinement effects that constrain the injected chain; and
ratcheting and pulling forces associated with DNA-binding proteins in
the host cell cytoplasm.

Considerable effort has been focused on the energetics of packaging
and ejecting DNA in phage. In particular, theoretical
work~\citep{riemer78, odijk98, kindt01, purohit03, tzlil03, odijk04,
metzler04, purohit05} has shown that the dominant source of stress on
the DNA in the capsid results from strong repulsive interactions
between neighboring portions of double helix that are confined at
average interaxial spacings as small as 2.5 nm. Another major
contribution comes from the bending stress that arises from the capsid
radius being smaller than the DNA persistence length. The force needed
to package the genome against this resistance is provided by a virally
encoded motor protein that pushes in the DNA along its length. Recent
laser tweezer measurements~\citep{smith01} have confirmed that this
force increases progressively as packaging proceeds, i.e., as the
chain becomes more crowded and bent, reaching values as large as $50$
pN upon completion. Conversely, the force \textit{ejecting} the DNA
upon binding of the phage to its membrane receptor has been
shown~\citep{evilevitch03, grayson05} to \textit{decrease}
monotonically from tens of piconewtons to zero as crowding and bending
stress are progressively relieved. In the current paper we consider
the {\it dynamics} of phage ejection and attempt to distinguish the
relative importance of these large, varying, ``internal'' forces and
the binding particles in the external solution (bacterial cytoplasm).

It is useful at the outset to consider the simple diffusion limit of
the translocation process. More explicitly, consider the case in which
a chain is threaded through a hole in a membrane dividing one solution
from another. If the chain is free, i.e., in the absence of pushing or
pulling forces and of binding particles, it will simply diffuse along
its length, experiencing a friction associated with its passage
through the membrane and the viscosity of the solution. The time
required for its translocation from, say, the left to the right will
be $L^2/ 2D=\tau_{d}$, where $L$ is the length of the chain
and $D$ is its effective translational diffusion coefficient.

Suppose now that particles are added to the right-hand solution
which bind \emph{irreversibly} to the chain at regularly spaced
sites as soon as they diffuse into the solution. Then, if $s$ is
the spacing between these binding sites, the diffusion of the
chain will be \textit{ratcheted} each time another length $s$ has
entered the solution~\citep{simon92, peskin93}, corresponding to the fact
that the chain cannot move backwards through the hole at a site
where a particle is bound. Accordingly, the time it takes for the
entire chain to appear on the right is simply given by $s^2/ 2D$
-- the time required for diffusion between a pair of neighboring
binding sites -- times the total number of sites, $L/ s$. It
follows that the overall translocation time in the presence of
perfect ratcheting is reduced by a factor of $s/L$ over that for
free diffusion. When the binding of particles is
\textit{reversible} -- they do not remain bound indefinitely,
thereby allowing some sites to diffuse backwards through the hole
-- the translocation time is increased by a factor of $(1+2K)$
compared to perfect ratcheting, where $K$ is the ratio of ``off''
and ``on'' rates for particle binding~\citep{simon92, peskin93}. Finally,
note that the ideal ratcheting time of $Ls/2D$ corresponds to a
velocity of $2D/s$ and hence, by the Stokes-Einstein relation, to
a \textit{force} of $2k_{B}T/ s$ pulling the chain into the
particle-containing solution~\citep{zandi03}.

When the particle binding is reversible, however, it turns out
that there can be an additional correction to the ratcheting dynamics,
one that can significantly \textit{shorten} the translocation time
below $Ls/ 2D=\tau_{\rm{idealratchet}}$. This effect requires that
the dffusive motion of the chain is slow enough and is due to
the fact that the entropy of reversibly bound particles increases
when there is more chain for them to explore.  As a result, the
entropy is an increasing function of chain length available in the
right-hand solution. Indeed, in the limit of fully equilibrated
binding, the system is equivalent to a
one-dimensional Langmuir adsorption problem~\citep{zandi03,
  ambjornsson04}(P. G. de Gennes, personal communication) . More explicitly, the
$1D$ Langmuir pressure can be written in the form
$P_{1D}=(k_{B}T/s)\ln\{1+\exp[(\epsilon+\mu)/ k_{B}T]\}$, where
$\epsilon > 0$ is the energy lowering of the adsorbing particles
upon binding and $\mu$ is their chemical potential in solution.
Note that in the limit of large binding energy ($(\epsilon+\mu)/
k_{B}T\gg 1$) this pressure reduces simply to $(\epsilon+\mu)/ s$,
which -- \textit{because pressure is force in a $1D$ system} --
can be directly interpreted as the force pulling on the chain due
to the reversible binding of particles. Note further, in the large
binding energy limit, that this force is necessarily large
compared to the ideal ratcheting force, $2k_{B}T/s$~\citep{zandi03}.

Ambjornsson and Metzler~\citep{ambjornsson04} have recently clarified the various timescales
that determine the different regimes of chain translocation in the
presence of ``chaperones'', i.e., binding particles. The first,
$\tauo$, is the time needed for the chain to diffuse a distance of
order $s$, the separation between binding sites. The second and third
are $\tauocc$ and $\tauun$, the characteristic times that a binding
site remains occupied and unoccupied, respectively. $\tauocc$ and
$\tauun$ are related by the equilibrium relation,
\bea
{\tauocc \over \tauun} = \exp \left({\epsilon + \mu \over k_{B}T} \right),
\eea
Finally, $\tauun$ can be approximated by the typical
time it takes for a particle to diffuse a distance of order $R$ ($\simeq
c_0^{-1/3})$ between binding free particles:
\bea
\tauun = {R^2\over 2D_0} \simeq {1\over D_0c_0^{2/3}},
\eea
where $D_0$ is the diffusion coefficient of the particles. One can
then distinguish between three different regimes:
\begin{enumerate}
\item Diffusive regime: $\tauo \ll \tauun, \tauocc$. Here the binding
  particles are irrelevant to the chain translocation because the
  chain diffuses its full length in a time too short for the particles
  to bind.
\item Irreversible binding regime: $\tauun \ll \tauo \ll \tauocc$. Here
  particles bind essentially irreversibly on a time scale short
  compared to the time it takes for the chain to diffuse a distance
  between binding sites. We shall refer to this as the ``ratcheting''
  regime.
\item Reversible binding regime: $\tauun, \tauocc \ll \tauo$. Here
  diffusion of the chain along its length is slow compared to the time
  required for an ``on''/``off'' equilibrium of the binding particles
  to be achieved. We shall refer to this as the ``Langmuir'' regime.
\end{enumerate}

It is also important to clarify some relevant \textit{length} scales
involved in the problem. Specifically, we distinguish between two
extremes of how the separation, $s$, between binding sites compares
with the range, $\delta$, of the attractive interaction between binding
particle and chain. Pure and ``perfect'' ratcheting will arise when
$\tauun \ll \tauo \ll \tauocc$, independent of the relative values of
$\delta$ and $s$. ``Imperfect'' ratcheting will arise when $\tauun, \tauocc
\ll \tauo$, but $\delta \ll s$. The translocation time for the
``imperfect'' ratchet is higher than the ``perfect'' ratchet by a
factor of $(1+2K)$. Finally, when $\tauun, \tauocc \ll \tauo$ and $\delta
\approx s$, in addition to the ``imperfect'' ratchet we also have a
Langmuir force on the chain. Note that if the binding free energy
between DNA and the binding proteins is very large then $K \ll 1$, and
the imperfect ratchet is no different than the perfect one. In this
paper we will always take this limit.

Before proceeding further it is instructive to make some numerical
estimates. Within this simple
translocation model all time scales are naturally referenced to
that for pure translational diffusion of a chain along its length,
and hence to the diffusion coefficient $D$ introduced earlier. In
reality, however, the DNA ejection process is enormously more
complicated, since the chain moving through the tail of the phage
is feeling not only the friction associated with the few hydration
layers surrounding it but also the viscous effects arising from
interaction with the inner surface of the tail just nanometers
away. Furthermore, this chain portion is  connected to the lengths
of chain inside the capsid and outside in the cell cytoplasm. The
chain remaining inside the capsid moves by reptating through
neighboring portions of still-packaged chain and/or by overall
rotation of the packaged chain. All of these latter motions
involve viscous dissipation that is insufficiently
well-characterized to enable realistic estimates of diffusion time
scales, even though one can distinguish between different
dependence on chain length for each of these dynamical
processes~\citep{gabashvili92, odijk04}. As a result, in pursuing
the simple translocation picture as a model for overall phage
ejection kinetics, we resort to using an effective diffusion
coefficient $D$ to define the unit of time, $\tau_d=L^2/ 2D$.

\begin{figure}[htb]
\begin{center}
\includegraphics[]{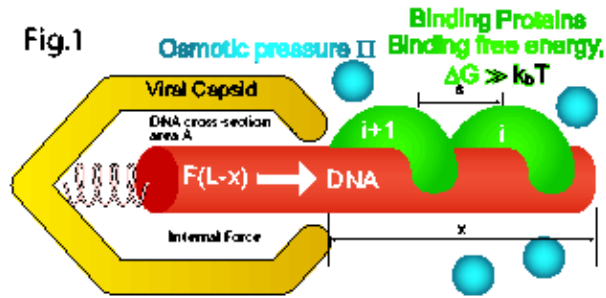}
\caption{\label{fig:schematic}Schematic showing the various
physical effects which assist bare diffusion in the process of
phage DNA ejection. The DNA cross-section is not shown to scale:
its diameter is $2-4$ nm, as compared with a capsid interior
diameter that is ten times larger. The spring denotes
schematically the stored energy density resulting in a force $F$
acting along the length $L-x$ of chain remaining in the capsid.
The small spheres denote particles giving rise  to an ``external''
(cytoplasmic) osmotic pressure $\Osmotic$, while the green
particles labeled $i$ and $i+1$ are successive binding
particles.}
\end{center}
\end{figure}

A strong upper bound for $D$ can be obtained by considering the
part of the dissipation arising as the chain moves through the
tail portion of the virus. Taking into account only the friction
between the DNA and the fluid in the tail we have, for
example~\citep{landau87, gabashvili92}, $\zeta=2 \pi l \eta/ \ln(\Delta/ d)$.
Here $\zeta$ is the friction coefficient, $l$ is the length of the
tail, $\eta$ is the viscosity of water, $\Delta$ is the inner
diameter of the tail, and $d$ is the diameter of the
double-stranded DNA. Taking $l=100\textrm{nm}$,
$\eta=10^{-9}\textrm{pN-s/nm}^2$, $\Delta= 4 \textrm{nm}$~\citep{tao98} and $d=
2\textrm{nm}$, we find $\zeta=9 \times 10^{-7} \textrm{pN-s/nm}$
and hence a diffusion coefficient ($D=k_{B}T/ \zeta$) of $5 \times
10^6 \textrm{nm}^2\textrm{/s} $. For a typical phage genome length
($L$) of $10 \mu\textrm{m}$, this in turn leads to a diffusional
translocation time ($\tau_{d}=L^2/ 2D$) of about $10$ seconds, not
unlike ejection times measured for phage {\ldphage}~\citep{novick88}.
Recall, however, that this estimate is based on a value for $D$
which is a strong \textit{upper} bound, because of all the viscous
dissipation contributions that were neglected, suggesting that the
actual unassisted diffusional time is likely several orders of
magnitude larger than this $10$ seconds estimate. Indeed, the
outcome of the work presented below is that  the translocation
time is shortened beyond $\tau_d$ by several orders of magnitude
by a combination of effects dominated by pressure in the capsid
and binding particles in the external solution.  A schematic of
the role of these various effects is shown in
Fig.~\ref{fig:schematic}.

The outline of our paper is as follows. In the next section we
include the effect of capsid pressure by formulating a
Fokker-Planck description of translocation driven by a combination
of diffusion and spatially varying force, i.e., a force pushing
the chain from one side that depends on the length of chain
remaining on that side (corresponding to the portion still in the
capsid and hence experiencing stress due to crowding and bending).
We evaluate the mean-first-passage-time (MFPT) for translocation
of an arbitrary length and thereby calculate the length ejected as
a function of time, using estimates of the spatially-varying
ejection force from recent theories of phage packaging energetics.
We find that the translocation times are 2 to 3 orders of
magnitude faster than the diffusional time. We also treat the case
of ejection into a volume comparable to the capsid size
(mimicking, say, studies in which phage are made to eject into
small vesicles that have been reconstituted with receptor protein
\citep{letellier99, bohm01}) and find that the ejection time decreases
from its maximum value when the confinement scale is made larger
\textit{or smaller} than the capsid size. In Sec.~\ref{sec:bp} we
treat the further speed-up in translocation due to ideal
ratcheting and to reversible particle binding, respectively. We
find that the simple ratcheting effect is small compared to that
arising from the entropic force of reversible particle binding.
The effect of reversible particle binding decreases the
translocation time by another order of magnitude beyond that due
to capsid pressure effects. We conclude in
Sec.~\ref{sec:discussion} with a discussion of related work by
others, of additional contributions to ejection dynamics that will
be studied in future theoretical work (in particular, the effect
of RNA polymerase acting on the ejected DNA), and of experiments
planned to test the various predictions made in the present work.

\section{\label{sec:force}Kinetics of ejection driven by packaging
    force} As discussed in Sec.~\ref{sec:intro}, we focus here on a
chain which has been confined in a viral capsid and which is ejected
from it through a hollow tail just big enough to accommodate its
diameter. To elucidate the essentials of this ejection process we
describe the translocation of the chain as a ``diffusion-in-a-field''
problem~\citep{sung96, lubensky99, muthukumar99}. In the present case,
involving the translocation of a linear polymer along its length, the
diffusion coordinate is a scalar, i.e., the length of chain $x$ that
has been ejected from the tail of the virus. The external field is
described by the potential energy $U(x)$ that gives rise to the force
$F(x)=-dU(x)/ dx$, pushing on the chain when a length $x$ of it has
been ejected. This force is due to the remaining chain length $L-x$
being confined inside the capsid and thereby subjected to strong
self-repulsion ($U_{\rm rep}$) and bending ($U_{\rm bend}$). The
corresponding potential $U(x)=U_{\rm rep}(L-x) + U_{\rm bend}(L-x)$ is
the free energy calculated in recent theories of DNA packaging in
viral capsids~\citep{tzlil03, purohit03, purohit05}. This energy is
seen to decrease dramatically as ejection proceeds (i.e., as $x$
increases), and so does the magnitude of its slope that constitutes
the driving force for ejection.

The one-dimensional dynamics of a diffusing particle in the
presence of an external field is a classic problem in stochastic
processes~\citep{kampen92}, and, as argued above, can be tailored
to treat the translocation of phage DNA under the action of an
ejection force $F(x)=-dU(x)/dx$. Accordingly, the probability
$p(x,t)$ of finding a length $x$ ejected at time $t$ is given by
the Fokker-Planck equation \bea {\partial p(x,t)\over
\partial t} = {\partial \over
\partial x}\left(D{\partial p(x,t)\over \partial x}+ {D\over
k_{B}T}{\partial U(x)\over \partial x}p(x,t)\right). \label{FPE}
\eea As part of this stochastic description of the
translocation-under-a-force process, it is natural to define a
mean-first-passage-time (MFPT), $t(x)$, that gives the average
time it takes for a length $x$ to be ejected in the presence of the
external field $U(x)$, namely~\citep{howard01}, \bea t(x) &=&{1\over
D} \int_{0}^{x}dx_1\exp\left(-{U(x_1)\over
k_{B}T}\right)\int_{x_1}^{x}dx_{2}\exp\left({U(x_{2})\over
k_{B}T}\right).\label{MFPT} \eea
 It is useful to consider several
limits of this general equation, the first corresponding to the
familiar case of no external field. From $U \equiv 0$ the
integrals in {MFPT} reduce trivially to $x^2/2D$, giving the
expected diffusion time, $t(x)=x^2/ 2D$.

For the case of \textit{constant} force, i.e., $U=-Fx+
\rm{constant}$, the integrals in {MFPT} can also be evaluated
analytically, giving~\citep{peskin93}
\bea
t_{\rm{ConstantForce}}(x)={x^2\over
D}{\exp[-\beta F x] +\beta F x
  -1\over (\beta Fx)^2}.\label{constF}
\eea
Here we have written $\beta$ for $1/k_{B}T$, and taken  $F =
-dU(x)/dx >0$ to denote the constant force \textit{driving}
translocation of the chain to the right. In Sec.~\ref{sec:bp} we
will apply Eq.~\ref{constF} \textit{locally}, over each
segment of length $s$ associated with a binding site, to calculate the ideal
ratcheting corrections to force-driven translocation. Note that
simple and ratcheted diffusion are overwhelmed by force-driven
translocation when $\beta FL \gg 1$ and $\beta Fs \gg 1 $,
respectively.

In the most general instance of spatially varying ``external''
field $U(x)$, as in the case of capsid-pressure-driven
translocation, the integrals in Eq.~\ref{MFPT} must be evaluated
numerically.    In this way we calculate $t(x)$ from
Eq.~\ref{MFPT} for the $U(x)$ determined from a recent
treatment~\citep{purohit03, purohit05} of the packaging energetics
in phage capsids. This provides a one-to-one correspondence
between each successive time $t(x)$ and the fraction of chain
ejected $x(t)/L$ at that instant.

\begin{figure}[htb]
\begin{center}
\includegraphics[]{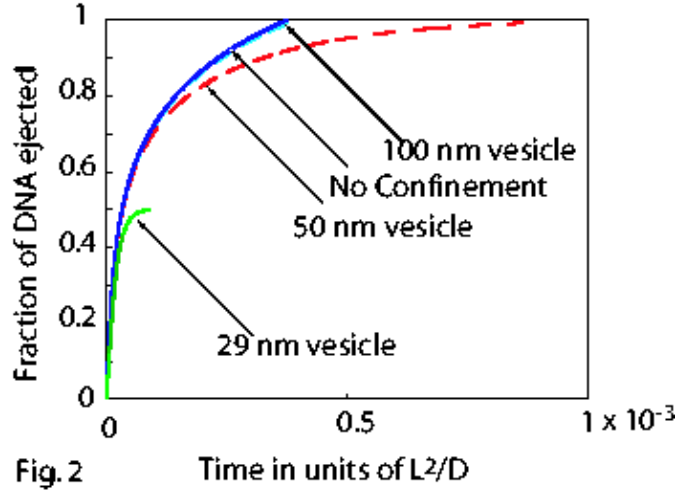}
\caption{\label{fig:vesicle}Ejection time for phage-\(\lambda\)
  injecting its genome into vesicles of radius 29, 50 and 100 nm. The
  capsid radius of the phage is 29 nm. It can be seen that the amount of
  DNA injection increases as the ratio of the vesicle radius to the
  capsid radius increases. On the time scale depicted here, there will
  be essentially no ejection due to pure diffusion (which takes place
  instead at times of order 1, in units of $L^2/D$). }
\end{center}
\end{figure}

 In Purohit \emph{et al.}~\citep{purohit03, purohit05} the DNA inside
 the phage capsid is assumed to be organized in a hexagonally packed
 inverse-spool. The potential $U(x)$ is expressed as a combination of
 the bending energy and the repulsive interaction between the DNA
 strands, and is given by,
\bea
\nonumber
U(x) &=& U_{\rm rep}(L-x) + U_{\rm bend}(L-x) \\\nonumber
&=& \sqrt{3} F_0 (L-x)(c^2 + c d)\exp(-d/c) \\
& &+ {2\pi k_b T\xi \over \sqrt{3}d}\int_{\Rin}^{\Rout}{N(r)\over r} dr. \,\label{packingenergy}
\eea
$F_0$ and $c$ are experimentally determined
constants~\citep{rau84} describing the interaction between neighboring
DNA strands, $\xi$ is the persistence length of DNA, $d$ is the
inter-strand spacing, $R_{\rm out}$ and $R_{\rm in}$ are the radius of
the capsid and the inner radius of the DNA spool, respectively, and
$N(r)$ is the number of hoops of DNA at a distance $r$ from the spool
axis. We are interested in finding the internal force on the phage
genome as a function of genome length inside the capsid. We do so
using Eq.~\ref{packingenergy} and simple geometrical constraints on
the phage genome inside the capsid. The number of loops $N(r)$ in
Eq.~\ref{packingenergy} is given by $z(r)/d$, where $z(r) = (R_{\rm
out}^2 - r^2)^{1/2}$ is the height of the capsid at distance $r$ from
the central axis of the DNA spool. The actual volume available for the
DNA -- $V(R_{\rm in}, R_{\rm out})$ -- can be related to the genome
length $L-x$ in the capsid, and the inter-strand spacing $d$, giving an expression for
$R_{\rm in}$ in terms of $d$, $R_{\rm out}$ and $L-x$. This relation
can be substituted for $\Rin$ in Eq.~\ref{packingenergy}, which then
can be minimized with respect to $d$ to give the equilibrium
inter-strand spacing as a function of the genome length $L-x$ inside
the capsid. In this way we determine the total packing energy as a
function of genome length inside the capsid $(L-x)$ or as a function
of the DNA length ejected $x$, i.e., $U(x)$. Using this result
and Eq.~\ref{MFPT} we can evaluate the MFPT, $t(x)$, for the DNA
ejection in {\ldphage} as a function of the length ejected. The
corresponding fraction ejected, $x(t)/L$, is shown as a function of
time in Fig.~\ref{fig:vesicle}, with the label ``no confinement''.

The value of $D$ can be estimated on the basis of this simple model by
the following procedure. The experiment by Novick and
Baldeschwieler~\citep{novick88} showed that in a buffer containing
$10$~mM of $\Mg$ it took roughly $50$ seconds for phage {\ldphage} to
completely eject its genome. The values for $F_0$ and $c$ in buffers
containing $\rm Mg^{2+}$ have been measured~\citep{rau84}. Since the
values measured for $5$mM and $25$mM $\rm Mg^{2+}$ were not
significantly different, we assume that the forces at $10$mM will be
identical, i.e., $F_0 = 12000~\rm{pN/nm^{2}}$ and $c =
0.3~\textrm{nm}$. Using these values in Eq.~\ref{MFPT} and numerically
evaluating it for $x = L = 48500 \times 0.34$~nm we find the total
time for {\ldphage} to eject its genome of $48.5$~kbp is $ t \approx
(10^{5} \rm{nm^2/D}) \textrm{seconds}$. Then, since this value is
experimentally estimated to be around $50$ seconds~\citep{novick88}, we
infer that $D \approx 10^{3} \rm nm^{2}/s$. This is about $3$ orders
of magnitude smaller than the $D$ estimated in Sec.~\ref{sec:intro},
consistent with all the sources of dissipation that were left out of
that estimate.


An interesting application of our estimates is to the experiments
in which viruses eject their DNA into lipid
vesicles~\citep{letellier04,novick88,bohm01,graff02}.  Here lipid
vesicles are reconstituted with the receptors recognized by the
phage of interest, and then mixed with a solution of the phage.
The phage binds to the receptor and ejects its DNA into the
vesicle. We argue that the amount of DNA ejected into the vesicle
and the corresponding time depend on the radius of the vesicle. In
particular if the vesicle has a radius comparable to that of the
viral capsid there will be a build-up of pressure inside the
vesicle due to the ejected DNA. Ultimately, the ejection process
will come to a halt when the force on the DNA from the capsid
equals the force from the vesicle side -- this can be thought of
similarly from the free energy perspective as a free-energy
minimizing configuration.  Hence, the ejection will not, in
general, be complete.

We can work out the ejection rate for this process as follows. If $x$
is the length of genome ejected into the vesicle, we denote the free
energies of the DNA inside the viral capsid and the vesicle by
$\Ucapsid(L-x)$ and $\Uvesicle(x)$, respectively. The total free
energy will be given by,
\bea U(x) = \Ucapsid(L-x) + \Uvesicle(x). \eea
As explained before, we already know $\Ucapsid(L-x)$ -- see
Eq.~\ref{packingenergy}; the expression for $\Uvesicle(x)$ can be
obtained similarly by assuming that the vesicle is like a
spherical capsid and the DNA configuration inside is similar to
that inside the viral capsid.  Our assumed structure for the DNA
in the vesicle is a highly idealized model, though we note that
electron microscopy on such vesicles demonstrates that DNA within
them can adapt highly ordered configurations~\citep{bohm01}. In the
limit where the vesicle radius is large compared to that of the
phage capsid we will recover the free injection result (DNA
ejecting from phage into the surrounding solution).

The injection process will stop when the total free energy reaches
a minimum, i.e., the total force on the DNA is zero. The time for
DNA injection is given by Eq.~\ref{MFPT}. We have worked out the
kinetics of the ejection for the bacteriophage \(\lambda\) (radius
$\approx 29$ nm) ejecting its genome into vesicles of radius \(29,
50, 100\) nm. The phage is taken to be suspended in a
solution of $\Mg$ ions, and similarly the vesicle, with
concentration that approximately gives the same values for $F_0$
and $c$ as discussed earlier. This yields a prediction for the
kinetics of injection for different vesicle radius. It can be seen
from Fig.~\ref{fig:vesicle} that when the size of the vesicle is
comparable to the capsid size there is only a partial ejection of
the DNA. When the vesicle size is almost twice the size of the
capsid nearly the entire genome is ejected, except for the last
part of the DNA, which takes ``extra'' time  because of the
resistance offered to it from the DNA inside the vesicle. Finally,
when the vesicle is more than three times the size of the capsid,
DNA gets completely ejected from the phage capsid as if there were
no vesicle. It is interesting to note that in the initial stages
of ejection all the curves for various vesicle sizes fall on one
another because there is no resistance to the injection, but as
the ejection proceeds each curve reflects a different resistance.

It is also possible that the arguments given above for {\it in
vitro} ejection into vesicles could be relevant to thinking about
ejection into the crowded environment of a bacterial cell
~\citep{zimmerman93, luby-phelps00}.  As a result of the crowding within the host
bacterium, the viral DNA may be subject to confinement effects
like those induced by vesicles.


\section{\label{sec:bp}DNA ejection in the presence of DNA
  binding proteins}

The \textit{E. coli} cell has as many as 250 types of DNA binding
 proteins~\citep{robison98}. Some fraction of these proteins likely
 bind either specifically or non-specifically to the phage genome as it
 enters the host bacterium. Accordingly, we consider what happens if the phage DNA is
  swarmed with binding proteins upon its entry into the host
 cell. Depending on the binding on/off rates, binding site density,
 and the strength of binding, we have a corresponding speed-up of the
 DNA injection into the bacterial cell. In this section we explore
 this effect and see how, in addition to the speed-up, it helps the
 phage inject its DNA against the osmotic pressure in the host cell.

Throughout the following analysis of particle binding effects, we
assume that the chain is stiff on length scales (e.g. $10$'s of
nanometers for double-stranded DNA genomes) large compared to the
size of the relevant binding particles (typically a few
nanometers). We also assume that the binding particles are
comparable in size to the distance between sites; for an estimate
of Langmuir forces in the more general case of larger binding
particles, see~\cite{ambjornsson04}.

\subsection{\label{sec:ratchet} DNA ejection due to the ratchet
 action.}

Consider a scenario in which host cell binding proteins
\emph{irreversibly} bind on to the DNA at a rate much faster than the
translocation rate. In such a case, once a binding site is inside the
cell, it is immediately occupied by a binding protein. If the protein
stays bound long enough, compared to the translocation time, it will
prevent thermal fluctuations from retracting the DNA back into the
capsid. As a result, the DNA will diffuse only between consecutive
binding sites, instead of along its complete length. Depending on the
spacing between the consecutive sites, it will bring about a
 speed-up in the translocation compared to when it is only
force-driven~\citep{peskin93}.

For simplicity we assume that the protein binding sites are
uniformly distributed along the length of the genome. If the
distance between the consecutive binding sites, $s$, is small
compared to genome length, i.e., $L \gg s$, we can assume that the
internal force on the genome due to the packaged DNA is effectively constant while the DNA
chain is diffusing between sites. In that case the MFPT, $t_i$,
for the DNA to translocate the distance $s$ between the binding
sites $i-1$ and $i$ is simply given by Eq.~\ref{constF}, with $x$
replaced by $s$, and $F$ replaced by $F_i$.  The internal force F
is of course a varying function of ejected length $x$, but to a
good approximation is constant over each interval of length $s$.
The subscript $i$ on the force F denotes this approximately
constant force on the DNA chain when the translocation is taking
place between the $i-1$ and $i^{\rm{th}}$ binding sites, i.e., when
length $(i-1)s$ has been ejected.

\begin{figure}[htb]
\begin{center}
\includegraphics[]{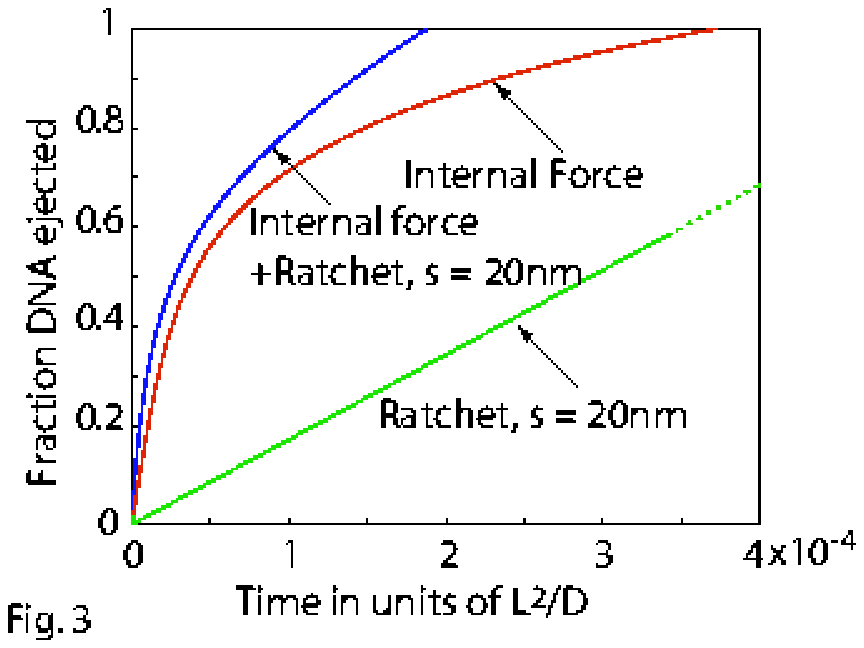}
\caption{\label{fig:disc}The fraction of DNA injected in phage $\lambda$
  as a function of time (in units of $L^2/D$) in the presence of binding
  particles that form a ratchet. The DNA injection purely due to the
  internal force is used as a benchmark, and the spacing between the
  binding sites $s = 20$ nm. It can be seen that the ratchet reduces the translocation time.
  The time required to internalize the genome solely by the ratcheting
  mechanism (see lower, straight line) is around twice the time taken
  for the purely internal force-driven mechanism.}
\end{center}
\end{figure}

The total translocation time for ejecting length $x$ of the DNA is
given by a sum over the MFPTs for all the sections of length $s$,
along the length $x$ ejected. The MFPT as a function of $x$ is
given by,
\bea
t(x)_{\textrm{Ratchet}+U(x)} = \sum_{i = 1}^{x/s}
t_{i}(F_i)\left|{\atop}\right._{\rm{Eq.~\ref{constF}}} \label{eq:time}
\eea
The corresponding plot for the fraction ejected, $x(t)/L$, as a
function of time is shown in Fig.~\ref{fig:disc} for $s = 20$ nm:
the ratcheting reduces the injection time by half as compared to
when the ejection results exclusively from the internal force. From
Eq.~\ref{constF} it can be seen that the time will decrease
exponentially as the spacing $s$ decreases. But, since $s$ can not
be smaller than the size of the binding proteins, which is of the
order of $10$~nm, we have a lower bound on $s$. The important
\textit{qualitative} consequence of the ratchet will be seen (see
Sec.~\ref{sec:osmoticp}) to be its helping with internalization of
the complete phage genome against osmotic pressure, when internal
force alone is insufficient to carry it out.

\subsection{\label{sec:revb} Reversible force from the binding
proteins}

\begin{figure}[htb]
\begin{center}
\includegraphics[]{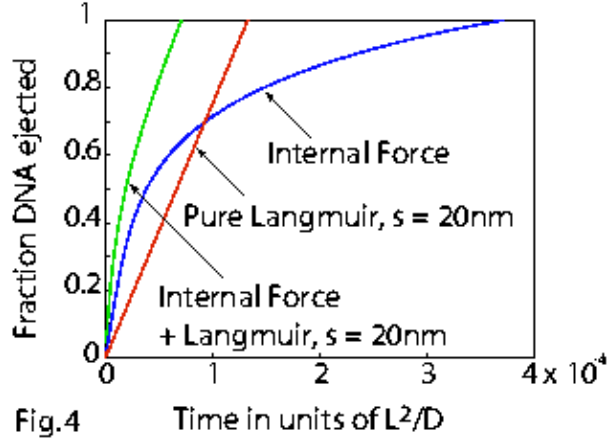}
\caption{\label{fig:cont} The fraction of DNA injected in phage
  $\lambda$ in the presence of  binding proteins that bind reversibly,  as a
  function of time (in units of $L^2/D$.) The presence of reversible
  binding proteins result in a pulling Langmuir force (see text). This
  pulling force significantly enhances the DNA ejection
  rate over that of the purely force-driven mechanism, by almost a
  factor of ten.}
\end{center}
\end{figure}

Consider another extreme scenario where DNA injects into a
reservoir of binding particles and the rate of translocation is
slow compared to the time required for the particles to bind and
unbind from the DNA. In this case, the binding proteins will come
to equilibrium with the DNA. As a result there will be an
adsorption force \emph{pulling} on the DNA, given
by~\citep{zandi03}(P. G. de Gennes, personal communication),
\bea
\nonumber F &=& {k_{B}T\over s}\ln\left\{1 +
\exp\left({\epsilon + \mu \over k_{B}T}\right)\right\}.\\ &\approx &
{\Delta G\over s}, \Delta G \gg k_{B}T
\eea
Here $\mu$ is the chemical potential maintained by the reservoir of
binding proteins, $\epsilon$ is the binding energy of the proteins
with the DNA, and $\epsilon +\mu = \Delta G(> 0)$ is the binding free
energy for the proteins. This adsorption force is the 1D Langmuir
pressure discussed in Sec.~\ref{sec:intro}. Now, since we have assumed
that $ \Delta G \gg k_{B}T$, the binding proteins are mostly
bound ``on'' the DNA and very rarely ``off''. So in addition to the
force there will also be a Brownian ratchet as discussed earlier. In
order to evaluate the MFPT we follow exactly the same process as in
the previous section with the addition of this Langmuir force $\Delta
G/s$ to $F_i$. The total MFPT is then given by Eq.~\ref{eq:time} with
$F_i$ replaced by $F_i + \Delta G/s$. We take a typical value of
non-specific DNA-protein binding free-energy of $\Delta G =
8k_{B}T$~\citep{shea85}. The plot corresponding to $s = 20$ nm is
shown in Fig.~\ref{fig:cont}. It can be seen that the Langmuir force
speeds up the genome translocation by almost an order of magnitude.
Not only that, but even if we do not have an internal force, this
mechanism (see Pure Langmuir) will inject the complete genome faster
than the internal force-driven mechanism. This is because after about
$50\%$ ejection, the internal force begins to drop below the constant
value of the Langmuir force. Indeed, from Fig.~\ref{fig:cont}, we see
that it is at an ejected fraction of about $0.5$ that the slope of the
``Internal Force'' curve drops below the constant slope (rate) of the
``Pure Langmuir'' plot.

The two cases we described are really two opposite extreme cases
for the DNA binding proteins. In reality the rate of binding and
the equilibration times may not be very fast (compared to
translocation times) and the translocation rates would lie
somewhere in between the rates evaluated in this section; for these
cases it is necessary to treat the dynamical coupling between
particle binding and chain diffusion~\citep{zandi03}.

\subsection{\label{sec:osmoticp} Binding proteins enable DNA ejection
against osmotic pressure}

Due to macromolecular crowding~\citep{zimmerman93}, the \emph{E. coli}
bacterium has internal osmotic pressures of about $3$
atm~\citep{neidhardt96}. The work of Evilevitch \emph{et al.}~\cite{evilevitch03} showed for
phage {\ldphage} that the ejection process can be partially/completely
inhibited by an application of osmotic pressure. Hence, it appears
that if the phage were to rely entirely on the driving force due to
the packaged DNA to eject its genome, the time scale for full ejection
would be prohibitively long. On the other hand, since we know that the
genome {\it is} completely internalized it seems likely that the
particle-binding mechanisms described above may play a key role in
{\it in vivo} DNA translocation. In this section we will see that the
task can be accomplished by the Brownian ratchet and the 1D Langmuir
force mechanism discussed in the preceding
Secs.~\ref{sec:ratchet}~\&~\ref{sec:revb}.

\begin{figure}[htb]
\begin{center}
\includegraphics[]{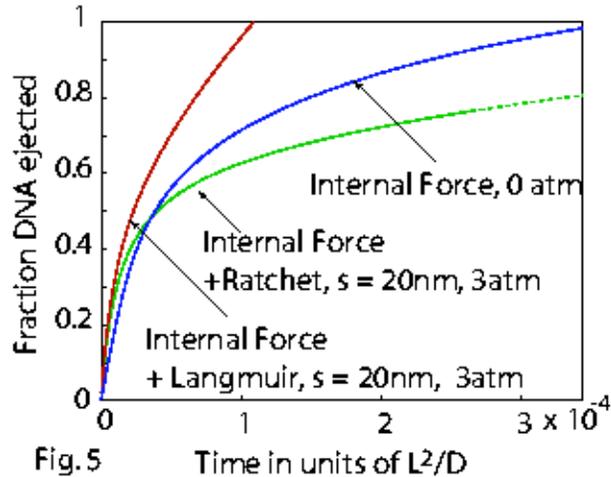}
\caption{\label{fig:osmotic}The fraction of DNA injected in phage
  {\ldphage} as a function of time (in units of $L^2/D$) for the case
  in which there is a resistive force due to osmotic pressure. We
  compare the roles of the Langmuir force and the ratchet effect in
  ejecting the phage DNA against osmotic pressure. The spacing $s$ is
  taken to be $20$nm and the osmotic pressure in the cell is around
  $3$atm. It can be seen that the Langmuir force easily pulls the DNA
  against this pressure. The DNA translocation by the Brownian ratchet
  requires a much longer time, but it still suceeds in pulling out the
  genome at time-scales not much longer than the ejection by
  internal force alone with zero osmotic pressure.}
\end{center}
\end{figure}

To see how the Brownian ratchet can internalize the genome against the
osmotic pressure, we use the following procedure. If the osmotic
pressure in the host cell is $\Osmotic$, the resisting force acting on
the DNA can be approximated~\citep{purohit05, tzlil03} by $\Fosmotic =
\Osmotic \pi \RDNA^2$, where $\RDNA$ is the radius of the DNA (about
$1$nm). For an osmotic pressure of $3$ atm the osmotic force is then
estimated to be around $1$ pN. We can now replace the term $F$ in
Eq.~\ref{constF} with $F_{i} - \Fosmotic$ to evaluate the MFPT, $t_i$
for the injection of the DNA segment between binding sites $i-1$ and
$i$. This time $t_i$ is then summed over all $i$, as in
Eq.~\ref{eq:time}, to give the time $t(x)$ and hence $x(t)/L$. This
fraction is plotted in Fig.~\ref{fig:osmotic} for the case of spacing
$s = 20$ nm between binding sites, and for an osmotic pressure of $3$
atm.

It can be seen from the figure (bottom curve) that the time
required for internalizing the genome is comparable to the time it
takes for phage to inject its genome purely by the internal force,
when there is no osmotic pressure.  The internal force for {\ldphage}
(data not shown) at around $50\%$ DNA ejection is approximately $1$
pN, i.e., of the order of $\Fosmotic$. It can be seen from
Fig.~\ref{fig:osmotic} that the slope of the curve showing
ejection in the presence of ratcheting and osmotic pressure starts
decreasing at that percentage of ejection. The average force
produced by a Brownian ratchet is $2k_{B}T/s \approx 0.4$pN for $s
= 20$nm~\citep{zandi03, peskin93}. At $60-70\%$ ejection the
internal force is around $0.5$pN; the total driving force is then
approximately $0.5 + 0.4 = 0.9$ pN, which is almost the same as
$\Fosmotic$. This force hence works to eject the genome against the
external osmotic force. When around $15\%$ of the genome is left
in the phage capsid, the internal force is almost zero. At this
point there is only a small amount of the genome still to be
ejected and a small differential of $\Fosmotic - F_{\rm ratchet}
\approx 0.5$ pN to be worked against. This is accomplished by the
Brownian motion of the DNA.

Now take the second case, when in addition to forming a ratchet
we have a 1D Langmuir pressure, as discussed in
Sec.~\ref{sec:revb}. To include the effect of the osmotic pressure
we have to subtract the osmotic force $\Fosmotic = \Osmotic \pi
\RDNA^2 $ from the driving force $F_i + \Delta G/s$ and use the
result in Eq.~\ref{constF}. This means that, so long as $(\Delta
G/s - \Osmotic \pi \RDNA^2) \ge 0$, we will always have DNA
ejection faster than or the same as that for the purely force
driven non-osmotic pressure case. For the numbers we took in the
preceding sections, $\Delta G/s = 8k_bT/s \approx 1.6 pN$, which
is greater than $\Fosmotic \approx 1 $pN. This implies that the
phage would inject its genome faster than in the purely
pressure-driven mechanism.

\section{\label{sec:discussion} Discussion and Conclusion}

This paper addresses the problem of the kinetics of phage injection
and the various mechanisms responsible for it. We make use of the
available experimental data, existing models for phage packaging, and
classical Fokker-Planck theory, to make predictions about
translocation rates for phage DNA ejection. The key quantitative predictions
described in this paper are:
\begin{itemize}
\item{\emph{Dependence of ejection rates on driving pressure.} As
  shown in Fig.~\ref{fig:vesicle}, the  driving force due to the
  packaged DNA speeds up the
  ejection process by $2-3$ orders of magnitude over free diffusion,
  and thus is a major contributor to the process of injection. Also,
  in the {\it in vitro} setting, the smaller the vesicle into which ejection occurs, the smaller the
  amount of DNA injected. In addition, for genomes of the same size,
  the time required for the ejection of the DNA is larger than when
  into a bigger vesicle.}
\item{\emph{Dependence of ejection rates on the presence of
  irreversible DNA-binding proteins.} Ratcheting enhances the DNA
  ejection rate from the viral capsid. The speed-up is minor when
  compared to internal force-driven ejection (see
  Fig.~\ref{fig:disc}), but as seen from Fig.~\ref{fig:osmotic} it is
  sufficient to pull out the genome against osmotic pressures of up to $3$ atm found inside the bacterial cell.}

\item{\emph{Dependence of ejection rates on the presence of reversible
  binding proteins.} The reversible binding of proteins exerts a 1D
  Langmuir force on the DNA. It can be seen from Fig.~\ref{fig:cont}
  that the presence of this phenomenon significantly enhances the DNA
  ejection rate beyond that due to pressure in the viral capsid. From
  Fig.~\ref{fig:osmotic} it is clear that this force is sufficient to
  efficiently internalize the phage genome against osmotic pressures of
  up to $3$ atm in the bacterium.}
\end{itemize}

We have several  biological examples in mind when we treat these
  ejection mechanisms. In bacteriophage T5 the DNA injection occurs in
  two steps. The first step transfer, which involves ejection of
  around $10\%$ of the phage genome, is driven by the internal
  force~\citep{letellier04}. There is then a brief pause, when a
  protein is synthesized that is implicated in the degradation of
  the host chromosome, thereby freeing the large number of proteins
  that had been bound to it. These latter proteins are now available
  for binding to the injected portion of the phage genome and for
  pulling the remaining DNA into the cell, via the ratcheting
 and Langmuir mechanisms.

 Similar ideas to those proposed here
 might also prove useful in those cases where the viral genome is
 translocated as a result of the binding of motor proteins which
 themselves translocate along the DNA.
 One such example is  the pulling force by the NTP-driven
RNA polymerase (RNAP).  RNAP is a very strong motor and can exert
forces of up to $14$ pN~\citep{wang98}. As described by Molineux
and coauthors~\citep{molineux01, kemp04} transcription by RNAP is
the major mechanism for DNA injection from wild-type T7 into
\emph{E. Coli} and is an intriguing additional active mechanism
that is of great interest to treat theoretically as well.  The
calculations presented here call for a more systematic
experimental analysis of the extent to which proteins bind onto
phage DNA as it enters the infected cell.

In this work we have analyzed various effects of DNA translocation
of internal capsid pressure and ``exterior'' (cytoplasmic) binding
proteins that can be tested by a variety of \textit{in vitro}
experiments involving phage ejection kinetics into synthetic
vesicles and through membranes formed over holes in planar
partitions. In these ways one can separately control the capsid
pressures (by varying salt concentrations or genome length, for
example) and the nature and concentration of DNA-binding proteins
inside the capsid or on the other side of the membrane. In
addition, it will be important to examine the role of these
various mechanisms in determining the kinetics of genome delivery
\textit{ in vivo}.

\textit{Acknowledgments.} We acknowledge helpful discussions with K. Dill,
C. Henley, P. Grayson, A. Grosberg, J. Kondev, C. Y. Kong, I.
Molineux, M. Muthukumar, P. Purohit, T. Squires, A. Voter, P. Serwer,
S. Casjens, H. Garcia, D. Reguera, P. Wiggins and R. Zandi. RP
acknowledges the support of the Keck Foundation, NSF grant numbers
CMS-0301657 and CMS-0404031 and the NIH Director's Pioneer Award grant
number DP1 OD000217. MMI acknowledges the support of , NIH grant
number R01 GM034993. WMG acknowledges the support of NSF grant no
CHE-0400363.

\bibliographystyle{unsrt}
\bibliography{dynamicspaper}

\end{document}